\pdfoutput=1
\documentclass[journal]{IEEEtran}
\usepackage{cite}
\usepackage{amsmath,amssymb}
\usepackage{graphicx}
\usepackage{booktabs}
\usepackage{array}
\usepackage{url}
\usepackage{tikz}
\usetikzlibrary{arrows.meta,positioning,shapes.geometric}

\newcommand{\Gv}{\ensuremath{C_v}}
\newcommand{\zkattach}{\texttt{zk-attach/v0}}

\begin{document}

\title{Zero-Knowledge Predicate Proofs Between AI Agents:\\
A Measured, Cross-Protocol Gateway and the Source-Integrity Gap}

\author{Ashok Subbabhatta Gopalakrishna%
\thanks{Manuscript received August 22, 2026.}%
\thanks{The author is an independent researcher (e-mail: sg.ashok@gmail.com).}%
}

\markboth{Preprint}{Subbabhatta Gopalakrishna: Zero-Knowledge Data Minimization for Multi-Agent AI Systems}

\maketitle

\begin{abstract}
Multi-agent AI platforms move quickly from staging to production, but the way agents establish trust remains rudimentary: an agent either transmits raw data to a peer or accepts that peer's natural-language self-report that a value complies with policy. The first over-shares; the second is unverifiable and is exactly the channel prompt injection attacks. Prevailing responses emphasise identity, visibility, and post-hoc detection, and recent proposals for cryptographically enforced agent policy have been evaluated in simulation rather than execution. We take provable data minimisation between agents from proposal to running system. In our Zero-Knowledge Proof Gateway, agents exchange proofs of governance-defined predicates over private data rather than the data itself, so exposure is prevented by design rather than detected afterwards; because no interoperability protocol can carry such a proof, we propose a slot and implement it on both MCP and Agent2Agent from one endpoint. A 32-bit threshold predicate proves in 6.2\,ms and verifies in 1.0\,ms with a 608-byte Bulletproofs proof on one commodity vCPU; eleven adversarial experiments and nineteen protocol checks pass; and the system is deployed to Kubernetes with empirically verified network isolation. Our case study proves a retail client order is within its limit without revealing the amount, instantiating the GDPR data-minimisation principle as an enforced technical measure of the kind EU law now names explicitly. We then address the limitation no comparable work resolves: a predicate proof binds a statement to a committed value, never to the system of record. We give a construction fusing an enclave attestation with the proof in both directions, so verifying one artifact certifies jointly that the predicate holds and that the value was read by a specific measured binary, and test it against a mock authority.
\end{abstract}


\IEEEpeerreviewmaketitle

\section{Introduction}
\IEEEPARstart{M}{ulti-agent} AI systems, in which planner agents delegate to specialist agents that invoke tools, and then transfer tasks to other agents across security boundaries, are being developed at a pace faster than the security architecture beneath them. The industry surveys from early 2026 quantify the gap: in one study of 919 security executives and practitioners, 88\% of organizations reported a confirmed or suspected AI-agent security incident within the prior year, while 82\% of them simultaneously expressed confidence that existing policies guard against unauthorized agent actions, and only about one in five organizations had runtime visibility at all into what their agents access~\cite{gravitee2026}. An independent survey of 300 enterprise leaders found 97\% expecting an agent-driven security or fraud incident within twelve months~\cite{arkose2026}. Agent fleets have roughly doubled in recent months while confidence in their security has risen faster than the controls that would justify it~\cite{gravitee2026}.

The risk underneath these numbers has been termed in the industry as the \emph{lethal trifecta}~\cite{willison2025trifecta,beurer2025patterns}: an agent that (i) has access to private data, (ii) is exposed to untrusted input, and (iii) can communicate externally or forward data onward can be exploited to exfiltrate data by successful prompt injections. Every inter-agent link in a pipeline compounds the trifecta, because today an agent establishes trust boundaries with peers in one of two ways: it ships the raw data (``here is the order information, check if its value is within the policy''), or it accepts a natural-language claim from the peer (or from itself) (``I verified; the value is within policy''). The first maximizes exposure; the second is unverifiable, non-deterministic, and provides no meaningful non-repudiation.

The security community's response operates from a different perspective. The OWASP Top~10 for Agentic Applications~\cite{owasp2026}, MITRE ATLAS~\cite{atlas}, and Zero Trust reference architectures for agents~\cite{anthropic2025zt,nist800207} focus on \emph{who} may act (identity, principle of least privilege), \emph{what happened} (visibility, audit logging), and \emph{detecting} bad actors or agents at runtime. These controls focus primarily on which principal may cross a boundary; they do not control \emph{what data} crosses the boundary when a permitted interaction occurs. A successfully authenticated, fully logged agent that sends a salary or a diagnosis to a peer that only needed a yes/no policy answer has still over-shared, and detection-based tooling will notice the share after the fact.

This paper proposes a complementary control: \emph{provable data minimization between agents}. Data minimisation is not merely a design preference in the European regime; it is a legal requirement under GDPR Article~5(1)(c), and Article~25 asks that it be realised through technical measures rather than assurances~\cite{gdpr}. Instead of data or text, agents exchange zero-knowledge proofs (ZKPs) of governance-defined predicates (range membership, set membership, threshold checks, and Boolean compositions) over their private data. The consumer agent learns exactly one bit (the predicate holds) plus a binding commitment and an audit handle; the sensitive value never crosses the boundary at all.

In this work, we make the following contributions:
\begin{itemize}
\item \textbf{An implemented and measured system, not a simulation.} Prior proposals for cryptographic agent-policy enforcement report proving costs drawn from calibrated distributions rather than execution~\cite{aegis}. Every figure here is measured from a single logged run of open-source code, reproducible in one command.
\item \textbf{A threat model for inter-agent data exposure vulnerability} (Section~\ref{sec:threat}) that separates the failure modes of natural-language self-report (prompt-injected claims, non-deterministic verification, and lack of non-repudiation) and maps the pipeline stages to exposure and mitigation.
\item \textbf{The Proof Gateway architecture} (Section~\ref{sec:arch}): a governance-owned predicate registry, a verifying gateway, prover sidecars, a hash-chained audit log, and a concrete protocol extension (\zkattach) that adds a proof slot to tool calls and handoffs, which no current agent protocol provides~\cite{mcp,a2a,openaitools}; we realise it on both MCP and Agent2Agent over a shared verification path, showing the carriage mechanism is protocol-agnostic rather than a property of any one protocol.
\item \textbf{A regulated case study under EU law} (Section~\ref{sec:case}): an execution agent proves a retail client's order notional satisfies a pre-trade cap without revealing it, instantiating the data-minimisation principle of GDPR Article~5(1)(c) and the data-protection-by-design obligation of Article~25~\cite{gdpr} as an enforced technical measure rather than a policy commitment, in the manner EU law now explicitly contemplates for predicate disclosure over personal data~\cite{eidas2}.
\item \textbf{A reference prototype and adversarial evaluation} (Section~\ref{sec:eval}): two interchangeable engines (an auditable Sigma-protocol baseline and Rust Bulletproofs~\cite{bunz2018,dalekbp}), microbenchmarks across bit-widths and aggregation levels, a latency-budget analysis, eleven adversarial experiments covering tampering, replay, rogue predicates, and forgery, and a Kubernetes deployment whose prover isolation was verified by observation rather than assumed from manifests. The implementation, the logged reproduction run, and all figures are released at \url{https://github.com/45h0kg/zk-proof-gateway}.
\item \textbf{Attestation-bound predicate proofs} (Section~\ref{sec:attest}): a construction that mutually binds an enclave attestation and a predicate proof, so neither can be substituted for the other, together with a governance-signed \texttt{prover\_measurement} predicate type that places prover identity under the same signing authority as the business rule it accompanies. We implement the construction and its six-step verification chain, wired into the live governed path on both protocol surfaces, and test it against a mock attestation authority. We regard source integrity, rather than proving cost, as the binding limitation of cryptographically enforced agent policy.
\end{itemize}
Section~\ref{sec:limits} states the limitations and open problems, including the explicit exclusion of verifiable LLM inference (zkML), Section~\ref{sec:related} places the work against the nearest prior art, and Section~\ref{sec:conclusion} is the conclusion.

\section{Problem Setting and Threat Model}\label{sec:threat}
\subsection{Trust Domains and Actors}
Our primary environment is a \emph{single-organization multi-agent system}: agents operated by one enterprise but spanning different trust boundaries (trading desk, risk, HR, procurement), where internal least-privilege policies apply between agents similar to how they apply between departments. A secondary environment, \emph{cross-organization} agent-to-agent interaction, gears up the motivation (the peer is another company) but adds identity-federation and legal machinery that is out of scope here; the proof layer is identical in both.

Actors: a \emph{prover agent} holding private data fetched from a source of truth; a \emph{relying (verifier) agent} that must establish a policy fact; a \emph{gateway} operated by the platform; a \emph{governance team} defining policy predicates; and an \emph{adversary} who may fully compromise the prover agent's LLM context (prompt injection), tamper with messages in transit, replay old messages, attempt to register weakened predicates, and edit logs after the fact. We assume the gateway and the governance signing keys are not compromised (the gateway is small, deterministic, and auditable, which is the reason the LLM is not part of the verification path), and we also assume standard cryptographic hardness assumptions (discrete logarithm; random-oracle Fiat--Shamir~\cite{fiatshamir}).

\subsection{Exposure Across the Pipeline}
Table~\ref{tab:threat} maps the stages of a contemporary agent pipeline to their current exposure and the proposed mitigation.

\begin{table*}[t]
\caption{Inter-agent data exposure across pipeline stages: current practice vs.\ proposed proof-gateway mitigation.}
\label{tab:threat}
\centering
\begin{tabular}{p{2.6cm}p{6.2cm}p{6.6cm}}
\toprule
\textbf{Pipeline stage} & \textbf{Current exposure} & \textbf{Proposed mitigation} \\
\midrule
Planner $\rightarrow$ sub-agent delegation & Full task context, including sensitive fields, copied into the sub-agent's prompt & Delegate the task with predicate proofs replacing sensitive fields; sub-agent receives policy facts, not values \\
Tool call & Arguments serialized in the clear into the tool schema; logged by middleware and providers & \zkattach{} attachment carries commitment + proof; sensitive argument omitted or replaced by a reference \\
Agent $\rightarrow$ agent handoff & Raw records or natural-language summaries cross the boundary & Gateway-verified predicate result + audit handle crosses; data does not \\
Memory / RAG store & Values persisted in shared vector stores or scratchpads readable by later, differently-privileged sessions & Persist commitments and proof outcomes; re-verification is possible without re-exposure \\
Audit layer & Logs replicate the sensitive values a second time (log-store becomes a shadow database) & Hash-chained entries bind commitment, proof hash, predicate id@version, agents, and result, with no values \\
\bottomrule
\end{tabular}
\end{table*}

\subsection{Why Proofs Rather Than LLM Self-Report}\label{sec:whynotllm}
The alternative to sending the data is today usually an LLM's own assertion. There are three failure modes that make such assertions unsuitable as a trust primitive, and each is directly addressed by verifiable proofs:

\subsubsection{Prompt injection compromises the claim} An agent whose context contains untrusted input can be induced to assert compliance falsely; the assertion channel and the attack channel are the same channel~\cite{willison2025trifecta,beurer2025patterns,owasp2026}. A ZKP moves the trust verification out of the LLM entirely: soundness holds against a \emph{fully malicious} prover, so a compromised agent can falsify the prose but cannot produce a verifiable proof of a false predicate (Section~\ref{sec:eval}, tests T1/T7).
\subsubsection{Non-determinism and hallucination on verification tasks} LLM outputs are probabilistic, vary across samples, and are not reliable on precise numeric checks; a verification check implemented as ``ask the model'' has no soundness parameter. Proof verification is a deterministic algorithm with soundness error negligible in the security parameter~\cite{gmr1985}.
\subsubsection{No non-repudiation or auditability} A natural-language claim binds no agent: it names no precise statement, no policy version, and can be disavowed. A proof is bound, through the Fiat--Shamir transcript, to a specific predicate identifier \emph{and version}, a request nonce, the reference of the action being authorized, and the two agent identities, and the audit log stores the commitment and proof hash; any interested party can later re-verify the exact claim that was accepted (tests T3/T4/T6).

\section{Preliminaries}\label{sec:background}
\subsection{Zero-Knowledge Proofs}
An interactive proof system for a language $L$ between prover $P$ and verifier $V$ satisfies~\cite{gmr1985}: \emph{completeness} (for $x\in L$, honest $P$ convinces $V$ except with negligible probability); \emph{soundness} (for $x\notin L$, no cheating prover convinces $V$ except with negligible probability); and \emph{zero-knowledge} (for $x\in L$, every verifier's view can be simulated without the witness, so the interaction reveals nothing beyond the truth of the statement). Goldreich, Micali, and Wigderson showed every language in NP admits a zero-knowledge proof~\cite{gmw1991}, which grounds our use of Boolean \emph{compositions} of atomic predicates.

\subsection{Sigma Protocols and Fiat--Shamir}
A Sigma protocol is a three-move (commit--challenge--respond) proof of knowledge; Schnorr's identification protocol~\cite{schnorr1991} is the canonical instance, proving knowledge of $x$ with $y=g^x$ via $(a{=}g^w,\; e,\; z{=}w+ex)$ and check $g^z = a\,y^e$. The Fiat--Shamir transform~\cite{fiatshamir} replaces the verifier's random challenge with a hash of the transcript, yielding non-interactive proofs in the random-oracle model; the hash input must bind the \emph{entire} statement and context to avoid malleability~\cite{bernhard2012}. OR-composition of Sigma protocols~\cite{cds1994} proves that at least one of two statements holds without revealing which; this is the building block of the bit proofs proposed. A small-number Schnorr transcript is given in Appendix~\ref{app:schnorr}. \subsection{Pedersen Commitments}
A Pedersen commitment~\cite{pedersen1991} to $v$ with blinding $r$ over a group of prime order $q$ with independent generators $G,H$ is $C = vG + rH$. It is \emph{perfectly hiding} (for any $v$, $C$ is uniform over the group), \emph{computationally binding} under the discrete-log assumption, and \emph{additively homomorphic}: $C_1 + C_2$ commits to $v_1{+}v_2$. Homomorphism is what allows a verifier to derive a commitment to $\mathrm{cap}-v$ from a public cap and \Gv{} without learning $v$, which is the basis of the threshold predicate here.

\subsection{Range Proofs and Bulletproofs}
A range proof shows a committed $v$ lies in $[0,2^n)$. Bulletproofs~\cite{bunz2018} achieve proof size $2\lceil\log_2 n\rceil + 9$ group/field elements, logarithmic in $n$, with no trusted setup, under the discrete-log assumption, made non-interactive via Fiat--Shamir. The original work reports a 688-byte single 64-bit proof (secp256k1 encoding) verifying in 3.9\,ms, with batch verification amortizing to 470\,$\mu$s, and aggregation of $m$ proofs adding only $O(\log m)$ elements~\cite{bunz2018}; the dalek ristretto255 implementation realizes $32(9+2\log_2 n)$ bytes, i.e., 672 bytes at 64 bits~\cite{dalekbp}. Bulletproofs+ tightens the 64-bit proof to 576 bytes with comparable computation~\cite{bpplus}. We adopt Bulletproofs as the practical mechanism because financial predicates over 32--64-bit integer amounts are well matched to its performance profile.
\section{Method: The Proof Gateway}\label{sec:arch}
\subsection{Components}
Fig.~\ref{fig:arch} shows the architecture. Five components cooperate:

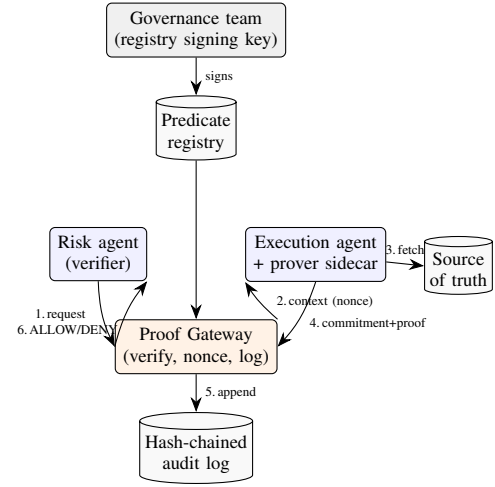
\begin{figure}[t]
\centering
\begin{tikzpicture}[
  box/.style={draw, rounded corners=2pt, align=center, font=\scriptsize, minimum height=7mm, inner sep=3pt},
  db/.style={draw, cylinder, shape border rotate=90, aspect=0.15, align=center, font=\scriptsize, inner sep=2pt},
  arr/.style={-{Stealth[length=2mm]}, font=\tiny},
  node distance=8mm and 6mm]
\node[box, fill=gray!12] (gov) {Governance team\\(registry signing key)};
\node[db, below=5mm of gov, fill=gray!5] (reg) {Predicate\\registry};
\node[box, below left=9mm and 3mm of reg, fill=blue!6] (ra) {Risk agent\\(verifier)};
\node[box, below right=9mm and 3mm of reg, fill=blue!6] (ea) {Execution agent\\+ prover sidecar};
\node[box, below=21mm of reg, fill=orange!10] (gw) {Proof Gateway\\(verify, nonce, log)};
\node[db, right=5mm of ea, fill=gray!5, yshift=-2mm] (src) {Source\\of truth};
\node[db, below=5mm of gw, fill=gray!5] (aud) {Hash-chained\\audit log};
\draw[arr] (gov) -- node[right]{signs} (reg);
\draw[arr] (reg) -- (gw);
\draw[arr] (ra.south) to[bend right=12] node[left, xshift=-1mm]{1.\,request} (gw.west);
\draw[arr] (gw.north east) to[bend left=8] node[right, xshift=1mm]{2.\,context (nonce)} (ea.south west);
\draw[arr] (ea) -- node[above]{3.\,fetch} (src);
\draw[arr] (ea.south) to[bend left=12] node[right, yshift=-1mm]{4.\,commitment+proof} (gw.east);
\draw[arr] (gw.west) to[bend right=-12] node[below left, yshift=-0.5mm]{6.\,ALLOW/DENY} (ra.south east);
\draw[arr] (gw) -- node[right]{5.\,append} (aud);
\end{tikzpicture}
\caption{Proof Gateway architecture. Sensitive values remain inside the prover's trust cell (right); only commitments and proofs cross agent boundaries.}
\label{fig:arch}
\end{figure}

\subsubsection{Predicate registry (governance-owned)} A catalog of predicate definitions, each comprising an identifier, version, type (\texttt{range\_leq}, set membership, Boolean composition), parameters (cap, bit-width, unit), and owner, and each signed by a governance key. This ownership decision is in itself a security control: if agents could define their predicates, a compromised agent would simply register a loose check (e.g., a cap of $2^{31}$) and comply with it. In the prototype the gateway verifies the governance signature both on publication and again on every read, so registry tampering is identified at use time (test T5).
\subsubsection{Proof gateway} A light, deterministic, LLM-free service that issues fresh request contexts (nonce, request id, predicate id@version, requester and prover identities, timestamp), verifies submitted proofs against registry-defined parameters, and appends audit entries. It is the only component trusted for verification, and it contains no language model.
\subsubsection{Prover sidecar} A proof engine located alongside the data holder, exposing \texttt{prove(predicate, context)}. Engines are plugged behind one interface; the prototype has a Sigma-protocol baseline and a Bulletproofs engine (Section~\ref{sec:eval}).
\subsubsection{Audit log} Append-only entries $h_i = \mathrm{SHA256}(h_{i-1}\,\|\,\mathrm{entry}_i)$ binding the commitment, proof hash, predicate id@version, agent identities, engine, and result. Verification is replayable forever; the sensitive value appears in no entry (test T6).
\subsubsection{Requesting and providing agents} LLM agents remain in the loop for orchestration but are removed from the trust path.

\subsection{Protocol Flow}
\begin{figure}[t]
\centering
\begin{tikzpicture}[font=\scriptsize, >={Stealth[length=2mm]},
  lifeline/.style={dashed, gray}]
\def\colA{0} \def\colB{2.55} \def\colC{5.1} \def\colD{7.3}
\node[draw, fill=blue!6, align=center, inner sep=2pt] (A) at (\colA,0) {Risk\\agent};
\node[draw, fill=orange!10, align=center, inner sep=2pt] (B) at (\colB,0) {Proof\\Gateway};
\node[draw, fill=blue!6, align=center, inner sep=2pt] (C) at (\colC,0) {Exec.\ agent\\(prover)};
\node[draw, fill=gray!10, align=center, inner sep=2pt] (D) at (\colD,0) {Source of\\truth};
\foreach \x in {\colA,\colB,\colC,\colD} \draw[lifeline] (\x,-0.45) -- (\x,-4.7);
\draw[->] (\colA,-0.8) -- node[above]{check(\texttt{cap@v1})} (\colB,-0.8);
\draw[->] (\colB,-1.35) -- node[above]{context \{nonce, ids, v1\}} (\colC,-1.35);
\draw[->] (\colC,-1.9) -- node[above]{fetch} (\colD,-1.9);
\draw[<-] (\colC,-2.3) -- node[above]{notional (private)} (\colD,-2.3);
\node[align=left, anchor=west] at (\colC+0.12,-2.85) {commit \Gv;\\prove $v\le$cap};
\draw[->] (\colC,-3.45) -- node[above]{envelope \{\Gv, $\pi$\}} (\colB,-3.45);
\node[align=left, anchor=west] at (\colB+0.1,-3.95) {verify; append audit};
\draw[->] (\colB,-4.45) -- node[above]{ALLOW + audit hash} (\colA,-4.45);
\end{tikzpicture}
\caption{Case-study sequence. The notional never crosses the dashed boundary between the prover's trust cell and the rest of the system.}
\label{fig:seq}
\end{figure}
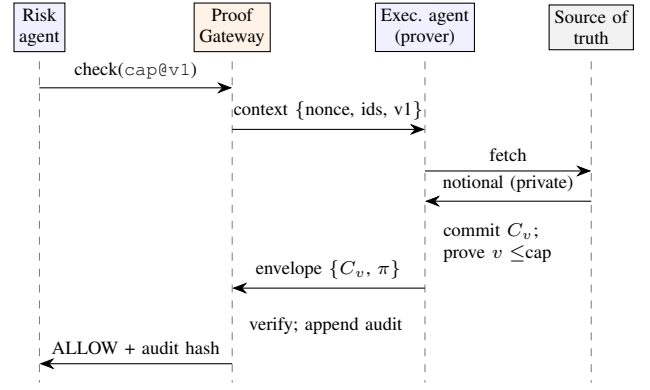

The end-to-end flow is: request $\rightarrow$ context issuance $\rightarrow$ source fetch $\rightarrow$ commitment $\rightarrow$ proof generation $\rightarrow$ verification $\rightarrow$ audit append $\rightarrow$ decision (Fig.~\ref{fig:seq}). Two properties are enforced by construction. \emph{Context binding}: the Fiat--Shamir transcript absorbs the full request context, including the reference of the specific action being authorized (e.g., the order identifier), before any commitment, so a proof verifies only for the exact request and action that solicited it; replay under a different nonce, action, agent pair, or predicate version fails (tests T3/T4). \emph{Freshness}: nonces are gateway-issued, so a prover cannot pre-compute proofs for future requests unless the deployment explicitly allows pre-staging (Section~\ref{sec:latency}).

\subsection{Source Integrity: What a ZKP Does and Does Not Bind}\label{sec:sourceint}
A range proof binds the statement to the \emph{committed} value; it cannot by itself show the committed value equals the system-of-record value. Two deployment topologies close this gap. \textbf{(A) Source-side prover (preferred):} the prover sidecar is alongside the source of truth; the co-location itself is the primary defense, as the orchestrator agent never holds the value, and the source furthermore signs the commitment to bind it to the system of record for transport. \textbf{(B) Agent-side prover with data attestation:} the source returns the value together with a signature over the commitment opening; not as secure (the agent can see the value locally) but deployable without source-system changes, and it still removes the value from all \emph{inter-agent} traffic, which is the scope of this paper. In either topology the guarantee also assumes the governed channel is the only path to the protected action: middleware must deny by default any governed action lacking a valid attachment, because no proof system constrains an endpoint that remains reachable without one.

\subsection{The \zkattach{} Protocol Extension}\label{sec:ext}
No mainstream agent-interoperability protocol carries a proof slot: the Model Context Protocol standardizes tool discovery and invocation over JSON-RPC and delegates authorization to OAuth-style flows~\cite{mcp}; Agent2Agent agent cards declare identity and authentication schemes for message parts~\cite{a2a}; function-calling schemas type tool arguments in the clear~\cite{openaitools}. All authenticate \emph{principals}; none can attest \emph{data properties}. We propose a backward-compatible attachment carried in MCP \texttt{tools/call} parameters or as an A2A message part, and implement both surfaces over a single endpoint (Section~\ref{sec:eval}):
\begin{figure}[h]
\centering
\begin{minipage}{0.95\columnwidth}
\scriptsize
\begin{verbatim}
"zk_attachment": {
  "schema": "zk-attach/v0",
  "predicate_id": "pretrade_notional_cap",
  "predicate_version": 1,
  "engine": "bulletproofs-ristretto255",
  "context": {"request_id": "...", "nonce": "...",
              "requester": "...", "prover": "...",
              "action_ref": "ord-9912", "ts": 0},
  "proof_b64": "..." }
\end{verbatim}
\end{minipage}
\caption{Proposed proof-attachment slot for tool calls and handoffs. The sensitive argument is omitted or replaced by a reference; absence of a required attachment for a registered predicate is deny-by-default. The attestation payload of Section~\ref{sec:attest} is an optional member of this same envelope, not a new schema version; see Section~\ref{sec:attest}-D.}
\label{fig:envelope}
\end{figure}

\subsection{Composition Across Multi-Hop Chains}\label{sec:composition}
Delegation chains raise the question of where verification happens. \textbf{(a) Chained per-hop proofs:} each hop requests and verifies its own proof; latency adds per hop, but the audit chain records every boundary at full granularity. \textbf{(b) Aggregated proof at a trust boundary:} $m$ range statements are proved in a single Bulletproofs aggregate whose size grows only $O(\log m)$~\cite{bunz2018}; one verification clears the boundary, at the cost of coarser audit granularity and prover-side work that grows linearly in $m$. Section~\ref{sec:eval} quantifies the tradeoff: at $m{=}16$, one 928-byte aggregate replaces 10{,}752 bytes of individual proofs and cuts total verification time roughly in half, while proving cost ($\sim$173\,ms on the test vCPU) argues for asynchronous aggregation rather than in-loop use.

\subsection{Scalability and Decentralized Verification}\label{sec:scale}
The gateway is logically central because policy definition and audit belong in one governed place, but verification itself is stateless and embarrassingly parallel: any replica holding the governance public key can verify any proof, so the verification role can be scaled horizontally or pushed into verifier-side sidecars that are as deterministic and model-free as the gateway itself. Decentralizing verification is sound only if audit completeness is preserved, for example through signed verification receipts returned to a shared append-only log; otherwise a compromised relying party could verify, act, and never log. The audit chain head is the one serialization point, addressable with per-domain chains or a transparency-log design. Policy definition remains centralized by design, since that centralization is the governance control of Section IV-A, not a scalability accident.

\section{Instantiation: Client Limit Compliance Without Disclosure}\label{sec:case}
\subsection{Scenario and Regulatory Grounding}
A trading-execution agent prepares an order for a retail client. Before submission, a risk agent must establish that the order notional does not exceed a per-client cap. Such client-level limits are ordinary risk and client-protection practice; the point of interest here is not that the limit exists but what enforcing it costs in disclosed personal data.

The regulatory interest for this paper, however, is not the control itself but the data that crosses the boundary while it is enforced. A retail client's order notional is personal data: it relates to an identified natural person. GDPR Article~5(1)(c) requires that personal data be adequate, relevant, and limited to what is necessary for the purpose, and Article~25 requires controllers to implement appropriate technical measures giving effect to data minimisation by design and by default~\cite{gdpr}. A risk agent that must establish only whether a threshold is respected does not need the amount; transmitting it is processing beyond what the purpose requires. The conventional implementation nonetheless sends the value and relies on organisational controls, which is a policy commitment rather than a technical measure. A predicate proof discharges the same control while making the minimisation property structural: the amount is not withheld by agreement, it is absent from the exchange.

That predicate proofs are the right instrument for this is not only our view. Regulation (EU) 2024/1183, which establishes the European Digital Identity Wallet, directs Member States to integrate privacy-preserving technologies, naming zero-knowledge proof, so that a relying party can validate whether a statement based on a person's data is true \emph{without revealing any data on which that statement is based}~\cite{eidas2}. That is a description in EU law of the mechanism this paper carries between agents; the same construction that answers ``is this person over eighteen'' without disclosing a birth date answers ``is this order within the client's limit'' without disclosing an amount. The predicate class is identical, and only the attribute changes.

Two further EU instruments bear on the audit component. GDPR Article~22 constrains solely automated decision-making affecting individuals, and the AI Act's record-keeping provisions for high-risk systems require that such systems technically permit automatic logging of events over their lifetime~\cite{aiact}. We note the scope precisely rather than overclaiming: the AI Act obliges logging capability, not cryptographic tamper-evidence, and the high-risk obligations phase in on a schedule that has been amended since adoption. The hash-chained audit here therefore exceeds the stated baseline rather than being compelled by it, and its value is that a decision can be re-verified later without retaining the personal data that justified it. In the concrete instance here the governance team publishes predicate \texttt{pretrade\_notional\_cap@v1}: $\mathrm{notional\_cents} \le 10^9$ (a \$10M cap) over 32-bit values; the private notional is \$7.35M.

\subsection{Predicate Construction}
The statement is
\begin{equation}
\mathrm{PoK}\{(v,r): \Gv = vG + rH \;\wedge\; 0 \le v \le \mathrm{cap}\}.
\end{equation}
Both engines reduce $v \le \mathrm{cap}$ to two range statements via the homomorphism: the verifier computes $C_d = \mathrm{cap}\cdot G - \Gv$ itself, and the prover shows $v \in [0,2^n)$ and $d = \mathrm{cap}-v \in [0,2^n)$. In the Sigma baseline, each bit $b_i$ of $v$ is committed as $C_i = b_iG + r_iH$ and proved to be 0 or 1 with a Cramer--Damg{\aa}rd--Schoenmakers OR-composition~\cite{cds1994} of two Schnorr proofs under a single Fiat--Shamir challenge; the value commitment is \emph{defined} as $\Gv = \sum_i 2^i C_i$, and the bit blindings of $d$ are constrained so $\sum_i 2^i C_{d,i} = C_d$ exactly, eliminating any separate linking proof. In the Bulletproofs engine the same two statements are proved directly (or as one 2-aggregate). In both, the transcript absorbs the request context before commitments, binding the proof to this predicate version, nonce, order reference, and agent pair.

\subsection{What Crosses the Boundary}
The wire envelope contains the schema tag, predicate id and version, engine identifier, the request context, a Pedersen commitment, and the proof (608 bytes in the Bulletproofs engine). The notional, the order, and any position data appear nowhere in the envelope or in the audit entry; the risk agent receives \texttt{ALLOW} and an audit handle. In the negative scenario, in which an agent's notional is \$12.5M, the honest prover API refuses to generate a proof at all because the predicate is false, and Section~\ref{sec:eval} shows that a \emph{malicious} prover cannot succeed either.

\section{Experiments}\label{sec:eval}
\subsection{Experimental Setup}
We implemented the architecture end-to-end: a governance-signed registry, gateway, hash-chained audit log, the \zkattach{} envelope, and two engines: (E1) an auditable $\sim$300-line pure-Python Sigma-protocol engine (Pedersen bit decomposition over secp256k1, CDS OR-proofs, Fiat--Shamir), and (E2) the Rust dalek Bulletproofs implementation over ristretto255~\cite{dalekbp}. E1 exists for auditability and pedagogy; E2 is the practical path. All measurements below are from a single logged run on one virtualized vCPU (medians). The environment, as reported by the container, is a single-core Intel Xeon (2.10\,GHz reported clock), 4\,GB RAM, Linux kernel 6.18 (Ubuntu 24.04), Python 3.12, and rustc 1.75 with \texttt{opt-level=3, lto}; because the host is virtualized, the exposed CPU string is not a physical part number and absolute timings will vary by host. These are feasibility evidence, not production benchmarks, and are consistent with published figures~\cite{bunz2018,dalekbp,bpplus}. The full reproduction log ships with the reference implementation.

\subsection{Results}
\begin{table}[t]
\caption{Range-proof engines, single statement $v\in[0,2^n)$ (medians).}
\label{tab:micro}
\centering
\begin{tabular}{@{}llrrr@{}}
\toprule
\textbf{Engine} & $n$ & \textbf{Size} & \textbf{Prove} & \textbf{Verify} \\
\midrule
Bulletproofs (E2) & 8  & 480\,B & 1.9\,ms & 0.42\,ms \\
                  & 16 & 544\,B & 3.3\,ms & 0.62\,ms \\
                  & 32 & 608\,B & 6.2\,ms & 1.02\,ms \\
                  & 64 & 672\,B & 11.5\,ms & 1.78\,ms \\
\midrule
Sigma bit-OR (E1) & 8  & 7.1\,KB & 191\,ms & 194\,ms \\
                  & 16 & 14.1\,KB & 362\,ms & 376\,ms \\
                  & 32 & 28.3\,KB & 749\,ms & 766\,ms \\
                  & 64 & 56.5\,KB & 1.45\,s & 1.50\,s \\
\bottomrule
\end{tabular}
\end{table}

\begin{table}[t]
\caption{Aggregation (pattern (b)): $m$ 64-bit statements, one Bulletproofs proof, vs.\ $m$ separate proofs.}
\label{tab:agg}
\centering
\begin{tabular}{@{}rrrrr@{}}
\toprule
$m$ & \textbf{Agg.\ size} & \textbf{$m$ singles} & \textbf{Agg.\ verify} & \textbf{$m$ verifies} \\
\midrule
1  & 672\,B & 672\,B & 1.8\,ms & 1.8\,ms \\
2  & 736\,B & 1{,}344\,B & 2.9\,ms & 3.6\,ms \\
4  & 800\,B & 2{,}688\,B & 4.3\,ms & 7.1\,ms \\
8  & 864\,B & 5{,}376\,B & 7.4\,ms & 14.2\,ms \\
16 & 928\,B & 10{,}752\,B & 13.7\,ms & 28.5\,ms \\
\bottomrule
\end{tabular}
\end{table}

\begin{figure}[t]
\centering
\includegraphics[width=\columnwidth]{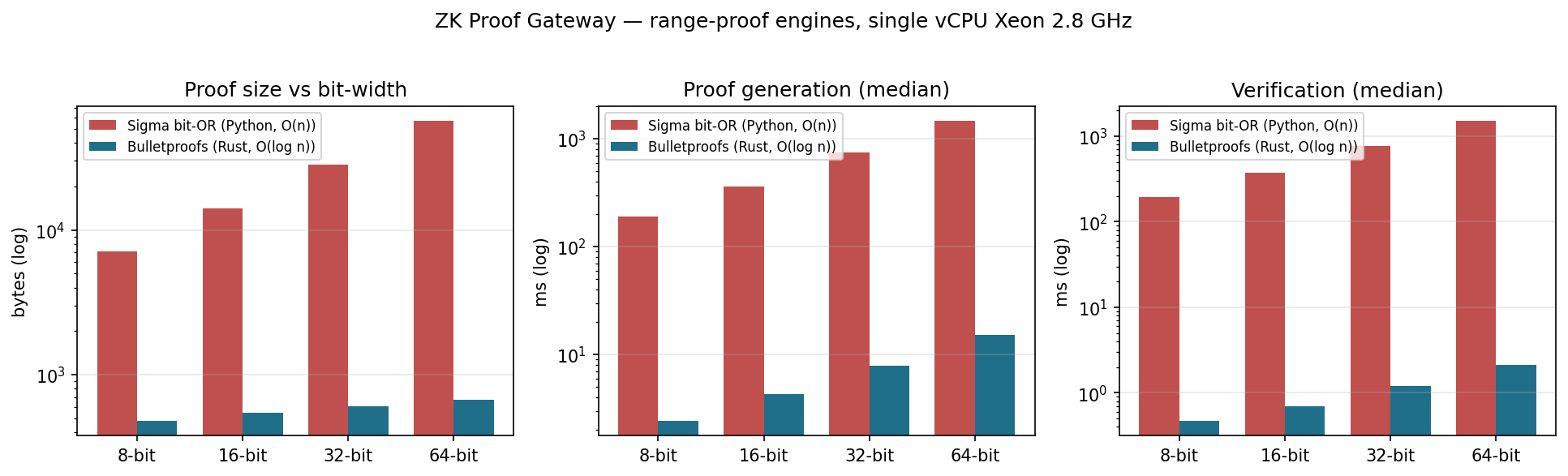}
\caption{Proof size, generation, and verification vs.\ bit-width for the two engines (log scale). The $O(n)$ vs.\ $O(\log n)$ separation is the practical argument for Bulletproofs at financial bit-widths.}
\label{fig:engines}
\end{figure}

\begin{figure}[t]
\centering
\includegraphics[width=\columnwidth]{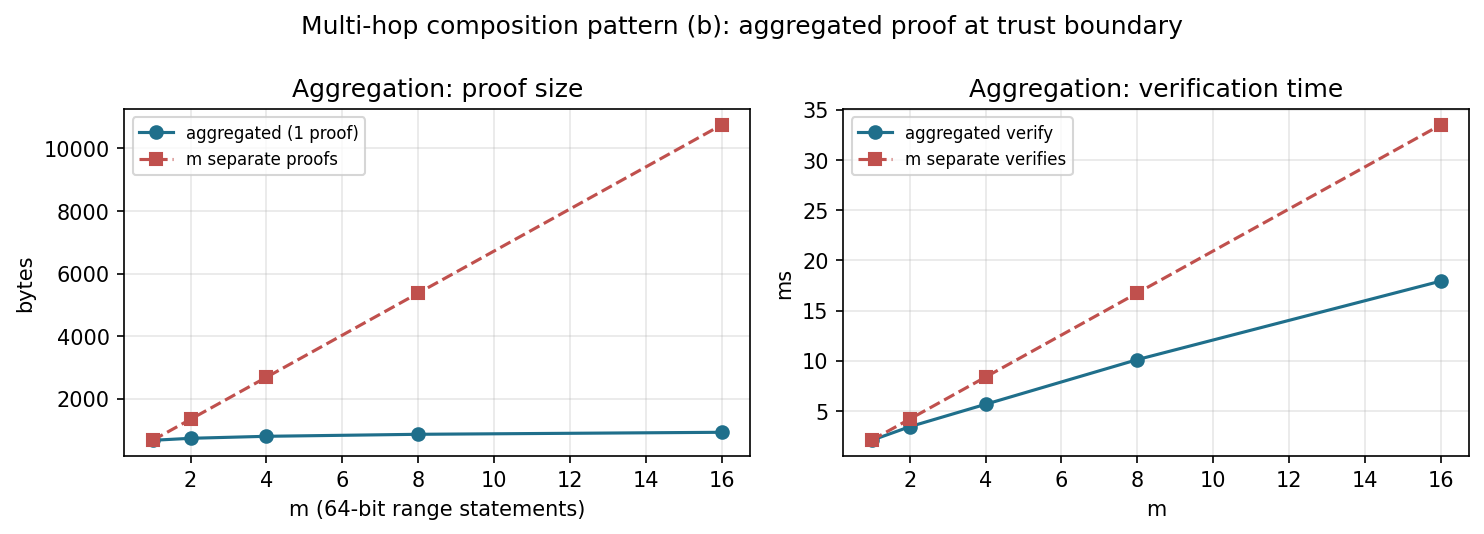}
\caption{Aggregated proofs at a trust boundary: size and verification vs.\ $m$ separate proofs.}
\label{fig:agg}
\end{figure}

Table~\ref{tab:micro} confirms the size asymptotics empirically: E2 grows by 64 bytes per doubling of $n$ (exactly $32(9+2\log_2 n)$ bytes, matching~\cite{dalekbp}), while E1 grows linearly. Table~\ref{tab:agg} quantifies composition pattern (b): at $m{=}16$ the aggregate is $11.6\times$ smaller and verifies about $2.1\times$ faster than sixteen singles, while prover time grows roughly linearly (173\,ms at $m{=}16$), supporting boundary-level aggregation computed asynchronously. Fig.~\ref{fig:engines} and Fig.~\ref{fig:agg} visualize both.

\subsection{End-to-End Case Study and Latency Budget}\label{sec:latency}
Running the Section~\ref{sec:case} scenario end-to-end (registry lookup with signature re-check, context issuance, prove, verify, audit append, in-process) yields p50 1{,}514\,ms / p95 1{,}538\,ms with E1; the same flow is bounded by $\approx$6\,ms prove + 1\,ms verify + I/O with E2. Mapped to pipeline locations: a real-time tool-call loop with a 10--50\,ms budget absorbs E2 comfortably, and proof generation can additionally be made asynchronous or pre-staged where the deployment permits nonce pre-issuance; agent handoffs with 100\,ms--1\,s budgets absorb even two-sided prove+verify with margin; audit-time re-verification is 1--2\,ms per entry. E1-class engines are prohibitive everywhere except offline audit and should be regarded as pedagogical.

\subsection{Adversarial Experiments}
Eleven experiments instantiate the Section~\ref{sec:threat} adversary; all behave as required. T0: honest baseline verifies. T1: a prover holding \$12.5M ($>$cap) cannot obtain a proof from the honest API (the statement is false). T2: a single bit-flip in a response scalar, and separately a swap of two bit commitments, are both rejected. T3: replay of a valid proof under a different nonce fails (context binding). T4: replay under a different \emph{predicate version} fails; policy updates invalidate old proofs by construction. T5: registry publication of an unsigned predicate, and of a predicate signed by a non-governance key, are both rejected. T6: an after-the-fact edit of one audit entry breaks the hash chain and is detected. T7: a malicious prover substituting a non-binary ``bit'' commitment ($C = 2G + rH$, an overflow attempt) with 2{,}000 randomized forgery attempts at the OR-proof achieved 0 acceptances; the same tamper and cross-context tests against E2 are likewise rejected. Beyond the in-process suite, we exercise the full \zkattach{} attachment over a live JSON-RPC channel between a client execution agent and an execution-venue server, across \emph{two} agent-interoperability surfaces served by one endpoint. Five MCP scenarios (a compliant order, a call with no attachment, a tampered proof, a valid proof replayed against a different order, and an over-cap notional) confirm deny-by-default, single-use context binding, and audited outcomes end to end. The same governed action is then exercised over Agent2Agent: an Agent Card advertises it as a skill, \texttt{message/send} carries the attachment in a message data part, and an order admitted via A2A lands on the same ledger the MCP surface reports. Both surfaces share one context issuer, one verification chain, and one audit log, so the guarantee is a property of the gateway rather than of either protocol. Nineteen checks pass in total. T7 is an implementation sanity check, not a security proof: soundness rests on the cited constructions~\cite{cds1994,bunz2018,bernhard2012}.

\section{Attestation-Bound Predicate Proofs}\label{sec:attest}
Section~\ref{sec:arch}-C states the limitation squarely: a range proof binds a statement to the \emph{committed} value, and topology A narrows the gap only by co-locating the prover with the source of truth. Neither says how a remote verifier confirms \emph{which binary} produced the committed value. This section gives a construction that does, and reports its implementation status honestly.

\subsection{Mutual Binding}
The goal is that verifying a single artifact certifies two things jointly: that the predicate holds over a committed value, and that the value was read from the source of truth by a specific, measured binary running in an attested enclave. Prior designs that combine trusted execution with zero-knowledge proofs tend to carry an attestation \emph{alongside} a proof, leaving each independently substitutable. We bind them in both directions, and both are necessary:

\emph{Direction A, attestation commits to the proof.} The enclave populates the attestation document's \texttt{report\_data} field with $\mathrm{SHA\text{-}384}(C_v \,\|\, \mathit{predicate\_id} \,\|\, \mathit{version} \,\|\, \mathit{nonce} \,\|\, \mathit{action\_ref})$. A valid attestation therefore cannot be replayed alongside a different commitment, predicate, or request.

\emph{Direction B, proof commits to the attestation.} Before any commitment is generated, the Merlin transcript absorbs $\mathrm{SHA\text{-}256}(\mathit{attestation\_document})$, extending the same context-binding step of Section~\ref{sec:arch}-B with a second, attestation-keyed message. A valid proof therefore cannot be presented under a substituted or unattested attestation.

Fiat--Shamir binds forward, from transcript to proof; \texttt{report\_data} binds backward, from the enclave's own output to the commitment it vouches for. The two directions close the same loop from opposite ends, which is why neither alone suffices.

\begin{figure}[t]
\centering
\begin{tikzpicture}[font=\scriptsize, scale=0.88, transform shape, >={Stealth[length=2mm]},
  lifeline/.style={dashed, gray}]
\def\colA{0} \def\colB{2.4} \def\colC{5.3} \def\colD{7.2}
\node[draw, fill=blue!6, align=center, inner sep=2pt] (A) at (\colA,0) {Relying\\agent};
\node[draw, fill=orange!10, align=center, inner sep=2pt] (B) at (\colB,0) {Gateway};
\node[draw, fill=green!8, align=center, inner sep=2pt] (C) at (\colC,0) {Enclave\\prover};
\node[draw, fill=gray!10, align=center, inner sep=2pt] (D) at (\colD,0) {Source};
\foreach \x in {\colA,\colB,\colC,\colD} \draw[lifeline] (\x,-0.45) -- (\x,-5.0);
\draw[->] (\colA,-0.8) -- node[above]{check(pred@v, action)} (\colB,-0.8);
\draw[->] (\colB,-1.3) -- node[above]{context $\{$nonce, action\_ref$\}$} (\colC,-1.3);
\draw[->] (\colC,-1.8) -- node[above]{read} (\colD,-1.8);
\draw[<-] (\colC,-2.2) -- node[above]{$v$ (private)} (\colD,-2.2);
\node[align=left, anchor=west, font=\tiny] at (\colC+0.15,-2.75)
  {commit $C_v$; attest with\\ \texttt{report\_data}$=$H$(C_v\|$pred$\|$v$\|$nonce$\|$action$)$\\ prove with transcript $\gets$ H(att.)};
\draw[->] (\colC,-3.6) -- node[above]{$\{C_v,\pi,\mathrm{att}\}$} (\colB,-3.6);
\node[align=left, anchor=west, font=\tiny] at (\colB+0.1,-4.1)
  {chain $\to$ root; measurement vs\\ registry; freshness; \texttt{report\_data};\\ verify $\pi$; audit};
\draw[->] (\colB,-4.8) -- node[above]{ALLOW $+$ audit hash} (\colA,-4.8);
\end{tikzpicture}
\caption{Attestation-bound flow. One gateway-issued nonce serves both the proof context and the enclave attestation call; \texttt{report\_data} binds the attestation to the commitment while the transcript binds the proof to the attestation.}
\label{fig:attest}
\end{figure}
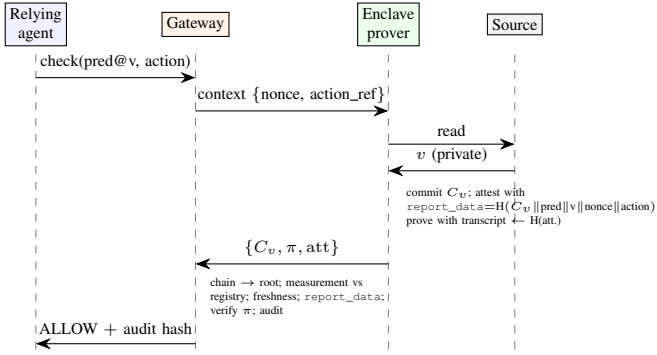

\subsection{Nonce Unification}
Fig.~\ref{fig:attest} shows the resulting flow. The substantive protocol requirement is not the hashing but the nonce. The \mbox{gateway-issued} nonce that already binds proof context must be the \emph{same} value passed to the enclave's attestation call, with identical expiry. Were the two allowed to diverge, an adversary could pair a fresh attestation with a stale but still-valid proof, or the reverse. Making one issued nonce serve both roles is what makes the two artifacts inseparable.

\subsection{Governance-Signed Prover Identity}
Expected measurements are not configuration. We add a \texttt{prover\_measurement} predicate type carrying the expected measurement, Schnorr-signed by the same governance key that signs threshold predicates and versioned the same way. The registry then states not only that the cap is a given value, but that the only binary permitted to assert it is the one with a given measurement. Prover identity becomes policy, under the same signing authority as the business rule it accompanies, and a compromised agent can no more substitute a prover than it can loosen a cap.

\subsection{Verification and Implementation Status}
The gateway's verification chain validates the attestation to its root of trust, checks the measurement against a registered \texttt{prover\_measurement} predicate where one exists, checks freshness against the issued nonce, recomputes and compares \texttt{report\_data}, verifies the proof under the attestation-bound transcript, and records the attestation digest in the audit entry alongside the existing fields.

\emph{Schema versioning.} The attestation travels as an optional member of the existing \zkattach{} envelope of Section~\ref{sec:arch}-D; it does not define a new schema version. An envelope either carries the attestation payload or does not, and the verifier selects the attested or unattested transcript accordingly, which keeps a v0 verifier that ignores the field interoperable with an attesting prover's unattested peers. Measurement \emph{policy} is a separate axis: whether a measurement is required is determined by the registry, not by the envelope. We keep these two axes independent deliberately, for reasons the implementation experience below makes concrete.

We implement this construction in the Rust engine as \texttt{attest-prove}, \texttt{attest-verify}, and \texttt{attest-measurement}, covered by 21 unit tests, and wire the verification chain into the live governed path on both the MCP and A2A surfaces, opt-in per deployment. \textbf{The attestation authority is currently a mock}: an Ed25519-signed document over module identity, timestamp, measurement, nonce, and user data, with a publicly derivable root seed standing in for a hardware root of trust. The construction, the binding, the verification chain, and the registry integration are real and tested; the hardware is not yet. Running this against AWS Nitro Enclaves and GCP Confidential Space is the remaining step, and Section~\ref{sec:limits} states what we expect it to cost and what it does not solve.

We are precise about what the mock does and does not establish, because the distinction matters for how this result should be read. What it does not establish is any security property: its root seed is publicly derivable, so it supplies no hardware root of trust, and it says nothing about an adversary who controls the platform. What it does establish is that the protocol is complete and functionally proven. The bidirectional transcript fusion produces artifacts that verify when and only when both bindings hold; the six-step verification chain executes in order against real signed documents and rejects at each step when its precondition fails; the governance-signed measurement predicate is parsed, matched, and enforced through the same registry path as a threshold predicate; and the whole sequence runs end to end on both protocol surfaces. Substituting a hardware attestation document for the mock changes the issuer of the document and the cost of validating it, not the state machine that consumes it. What is deferred is therefore the hardware timing evaluation and the platform trust root, not the design of the construction.

One implementation experience is worth recording because it is a property of the construction rather than of our code. An early version had the prover always attest while the gateway used the attested verification path only when a measurement predicate was registered. Because the two transcripts genuinely differ, every proof then failed to verify whenever no measurement policy existed. The defect was caught by rerunning the full end-to-end suite, not by unit tests, and the fix separates the two concerns: the transcript choice follows whether the envelope carries an attestation, while the measurement policy remains registry-driven and independent. Mutual binding makes attestation and proof inseparable by design, which means partial deployment is not a spectrum but a discontinuity.

\section{Limitations and Open Problems}\label{sec:limits}
\subsubsection{Only structured predicates reduce to circuits} Range, set-membership, threshold, and Boolean compositions cover a meaningful slice of enterprise policy, but fuzzy or judgment-based criteria (``is this communication client-appropriate?'') do not reduce cleanly to arithmetic statements. Wherever an LLM must judge, the architecture binds and audits the judgment's inputs and outputs, but it cannot make the judgment itself provable.
\subsubsection{No protocol slot exists yet} \zkattach{} is a proposal and not a standard. MCP, A2A, and function-calling schemas would each need a backward-compatible extension~\cite{mcp,a2a,openaitools}; we regard standardization as the most direct next step, and independent analyses of MCP have already identified the absence of capability attestation as a specification gap~\cite{mcpsec}.
\subsubsection{Metadata leakage} ZKPs mask the witness, not the traffic: which predicate was invoked, by whom, how frequently, and when remains visible to the gateway and network observers, and can itself be of a sensitive nature (a burst of \texttt{pretrade\_notional\_cap} checks reveals trading activity). Padding, batching, and private information retrieval are orthogonal mitigations not evaluated here.
\subsubsection{Governance is a socio-technical open problem} Who defines predicates, how versions roll out as policy changes (our version-binding makes stale proofs fail closed, but migration windows need policy), and what is supposed to happen on proof failure (hard deny versus human escalation) are not resolved; this proposal supplies the mechanism, not the operating model.
\subsubsection{Attestation is implemented but not yet hardware-validated} Section~\ref{sec:attest} gives the construction and tests it against a mock authority. Substituting real hardware introduces three things we have not measured: certificate-chain validation cost at the gateway, the size shift from a 608-byte proof to an envelope dominated by a multi-kilobyte attestation document, and the operational question of measurement churn, since rebuilding the prover changes its measurement and therefore invalidates previously registered values. Until that is done, source integrity still rests on the deployment discipline of topology A or B described in Section~\ref{sec:sourceint}.

\subsubsection{Enclave input imposes a deployment requirement} The construction certifies which binary produced a commitment; it does not by itself certify what that binary was fed. AWS Nitro Enclaves have no filesystem and no network, only a vsock channel to the parent instance, so a prover inside an enclave reaches the source of truth through a proxy on a host this threat model treats as untrusted. The requirement this imposes is standard rather than novel: the enclave must terminate an end-to-end encrypted channel to the system of record \emph{inside} its own memory, tunnelling TLS over vsock so that the parent relays ciphertext it can neither read nor alter. Two conditions make this sound. The trust anchor used to authenticate the source must be part of the measured image, so that a parent cannot redirect the session to an endpoint of its choosing without changing the measurement the registry pins; and any client credential the source requires must be released to the enclave under attestation, so that possession of it is conditioned on the same measurement. Under those conditions the parent retains only the ability to delay or deny traffic, which is an availability concern rather than an integrity one, and the honest residual is that we have specified this requirement rather than measured it.
\subsubsection{Verifiable LLM inference is out of scope} Proving that a model's own reasoning or inference was computed correctly (zkML) remains far too expensive for agent loops: state-of-the-art proofs for a 13B-parameter LLM inference take approximately 15 minutes to generate~\cite{zkllm}, versus the 6.2\,ms predicate proofs above. We state this explicitly as future work; the claim is that \emph{data-plane} predicates are provable today even though the \emph{model} is not.

\section{Related Work}\label{sec:related}
Table~\ref{tab:related} summarises how this work differs from the closest systems. The concept of proving policy compliance over private agent data is not ours: Aegis~\cite{aegis} proposed it first. What distinguishes this work is that the cryptography is executed rather than modelled, that the carriage mechanism is demonstrated on two protocols, and that the source-integrity assumption is stated and characterised rather than left implicit.

\begin{table*}[t]
\caption{Closest systems for cryptographically enforced agent policy. ``Source integrity'' asks whether the construction binds the proved value to the system of record. $^{\dagger}$Validated against a mock attestation authority; hardware evaluation deferred (Section~\ref{sec:attest}-D).}
\label{tab:related}
\centering
\scriptsize
\setlength{\tabcolsep}{4.5pt}
\begin{tabular}{@{}lllllll@{}}
\toprule
\textbf{System} & \textbf{ZK proves} & \textbf{Setup} & \textbf{Evaluation} & \textbf{Carriage} & \textbf{Audit} & \textbf{Source integrity} \\
\midrule
Aegis~\cite{aegis} & policy compliance & transparent & simulated & none & none & not addressed \\
Dist.\ Sentinel~\cite{distsentinel} & graph policy & trusted (Groth16) & measured & none & none & not addressed \\
zk-MCP audit~\cite{zkaudit} & message metadata & trusted & measured & MCP, async & audit service & not addressed \\
NiyamAI~\cite{niyamai} & judge executed & trusted & measured & interception & none & not addressed \\
Proof of Exec.~\cite{poe} & (no ZK) & --- & measured & agent runtime & signed records & trusted recorder \\
\textbf{This work} & \textbf{value predicate} & \textbf{transparent} & \textbf{measured, logged} & \textbf{MCP $+$ A2A} & \textbf{hash-chained} & \textbf{attestation-bound}$^{\dagger}$ \\
\bottomrule
\end{tabular}
\end{table*}
\subsubsection{Zero-knowledge architecture in deployed systems} Minimal-disclosure cryptography is proven at scale outside agent systems: Zcash's shielded transactions~\cite{zerocash}, and zero-knowledge or minimal-knowledge designs in messaging and credential systems. Our contribution is not new cryptography, nor the first application of zero-knowledge proofs to agent policy compliance; it is an end-to-end, measured, and governance-complete realisation of that idea at the inter-agent link.
\subsubsection{Agentic-security mainstream} OWASP's agentic Top~10~\cite{owasp2026} and threat taxonomy~\cite{owaspthreats}, MITRE ATLAS's agent-focused techniques~\cite{atlas}, and Zero Trust agent architectures~\cite{anthropic2025zt,nist800207} target identity, privilege, visibility, and detection. We position the Proof Gateway as complementary: those frameworks decide \emph{whether} an interaction may occur and observe it; ours minimizes \emph{what data} the permitted interaction carries. Notably, insecure inter-agent communication and human-agent trust exploitation appear as named risks in the 2026 OWASP list~\cite{owasp2026}; predicate proofs address the data-exposure component of both.
\subsubsection{ZKPs applied to agent systems} Closest to this work is the Aegis Protocol~\cite{aegis}, which also proves policy compliance over private agent data with a transparent-setup proof system (Halo2); its evaluation, however, is a discrete-event simulation in which proof latencies are sampled from a calibrated log-normal distribution rather than measured, and it does not address predicate governance, replay or action binding, protocol carriage, or audit chaining. Wu and Gong~\cite{distsentinel} include zero-knowledge predicate proofs between source and requesting sidecars as an optional extension to a taint-propagation architecture, using Groth16 and therefore requiring a trusted setup. Katkar et al.~\cite{niyamai} gate tool calls on a zk-SNARK that an isolated judge model executed correctly, hiding model weights rather than payload data. A further neighbour is the zero-knowledge \emph{audit} system of Jing and Qi~\cite{zkaudit}, in which agents exchange full message content over unmodified MCP, generate zk-SNARK proofs asynchronously after the session, and submit them to a third-party audit service that verifies statements about message format, type counts, and token usage; privacy holds against the auditor, not against the communicating counterparty, and the construction relies on a trusted setup. Our contribution differs in layer, purpose, and timing: the sensitive value is never transmitted to the relying agent at all; a value-predicate proof travels \emph{in lieu of} the data, synchronously, at the moment of trust establishment, under a transparent (no trusted setup) proof system. Huang et al.~\cite{ztidentity} apply DIDs, verifiable credentials, and ZKPs to agent \emph{identity} and attribute-based access control, i.e., the identity layer, which is orthogonal to the data-payload layer here. Proof of Execution~\cite{poe} verifies that governed agent actions occurred as recorded, but does so with signed execution records produced by a recorder inside the trusted computing base rather than with zero-knowledge proofs, and therefore inherits a trusted-component assumption where we inherit a committed-value assumption. Designs that pair trusted execution with zero-knowledge proofs for agent systems generally carry the attestation as a layer beside the proof, leaving each independently substitutable; the construction of Section~\ref{sec:attest} instead binds them to one another in both directions, and places the expected prover measurement under governance signature rather than deployment configuration. Stantchev's integrative review of verifiable law for machine societies~\cite{compjuris} surveys object-capability security, verifiable computation, policy-as-code, and agentic payment protocols, and concludes that the gap in this area is architectural rather than technological: the components exist but the discipline composing them does not. This paper is best read as discharging part of that agenda with running code, and as reporting which assumption survives the exercise. General agent threat models~\cite{narajala2025threat} inform Section~\ref{sec:threat}.
\subsubsection{Verifiable machine learning} zkLLM~\cite{zkllm}, optimizing systems such as ZKML~\cite{zkmleurosys}, surveys of the field~\cite{zkmlsurvey}, and production toolchains (EZKL, RISC~Zero)~\cite{ezkl} pursue proofs of \emph{inference}. We deliberately exclude this: our proofs are over the data agents exchange, not over the models that exchange it.

\section*{Reproducibility Statement}
All results in this paper come from a single logged run of an open-source reference implementation, available at \url{https://github.com/45h0kg/zk-proof-gateway}, reproducible with one command. The environment, engine versions, and per-stage outputs are recorded in the run log that ships with the code. Proof sizes are deterministic and host-independent; absolute timings depend on hardware, and we report figures from one virtualized single-vCPU host as feasibility evidence rather than production benchmarks. The adversarial suite and the end-to-end agentic protocol checks are deterministic pass/fail tests included in the release. The Kubernetes deployment described in Section~\ref{sec:eval} was installed and exercised on a local cluster; network-policy isolation of the prover was confirmed by observing connections fail with the policy applied and succeed with it withdrawn, rather than assumed from the manifest.

\section{Conclusion}\label{sec:conclusion}
Mainstream agentic security in 2026 is detection-based: it authenticates the agent, observes its behavior, and attempts to catch misuse. This paper argued for, designed, and prototyped a structurally preventive complement: when the sensitive value is not required for trust establishment, the interaction that establishes trust cannot expose it. The Proof Gateway replaces raw data and natural-language self-report with governance-defined predicate proofs; a compromised agent can still lie in prose, but it cannot produce a verifying proof of a false statement, cannot replay an old proof against a new request or policy version, and cannot register itself a weaker policy. Our reference prototype shows the costs are already compatible with real-time agent loops for financial bit-widths (608-byte proofs, single-digit-millisecond generation, millisecond verification), and our eleven adversarial experiments behave as the threat model requires. Next steps are: running the attestation construction of Section~\ref{sec:attest} against real Nitro Enclave and Confidential Space hardware and measuring what it costs, together with an implementation of the TLS-over-vsock input path and attestation-gated credential release that Section~\ref{sec:limits} specifies; a standardization proposal for the \zkattach{} slot in MCP and A2A, benchmarking on a broader catalog of enterprise predicates including set membership and Boolean compositions, and, as longer-horizon future work, revisiting verifiable inference as zkML costs fall.

\appendices
\section{Worked Schnorr Identification Example}\label{app:schnorr}
Group: the order-11 subgroup of $\mathbb{Z}_{23}^{*}$ generated by $g=2$ ($2^{11} \equiv 1 \bmod 23$). Secret $x=7$; public $y = g^x \bmod p = 2^7 \bmod 23 = 13$.
\begin{enumerate}
\item Prover picks nonce $w=3$, sends $a = g^w \bmod p = 8$.
\item Verifier sends random challenge $e = 5$.
\item Prover sends $z = w + ex \bmod q = 3 + 35 \bmod 11 = 5$.
\item Verifier checks $g^z \stackrel{?}{=} a\cdot y^e \bmod p$: $2^5 = 32 \equiv 9$, and $8 \cdot 13^5 \equiv 8 \cdot 4 \equiv 9 \pmod{23}$. Accept.
\end{enumerate}
Zero-knowledge: the tuple $(a,e,z)$ is simulatable without $x$ by sampling $z,e$ first and setting $a = g^z y^{-e}$; Fiat--Shamir sets $e = \mathrm{Hash}(\text{transcript})$ to remove interaction~\cite{fiatshamir}. Our bit proofs are OR-compositions of exactly this template~\cite{cds1994}.


\end{document}